\documentclass{ckm}                 

\usepackage{txfonts}            
\usepackage{amssymb}

\confname{Workshop on the CKM Unitarity Triangle, IPPP Durham, April
  2003}

\title{New Results from the PIBETA Experiment}

\author{D Po\v{c}ani\'c (for the PIBETA Collaboration)}
\address{Department of Physics, University of Virginia,
Charlottesville, VA 22904-4714, USA}

\begin{document}

\begin{abstract}We report interim results of the PIBETA experiment
analysis for the pion beta decay and pion radiative decay.  The former
is in excellent agreement with the SM predictions at the 1\,\%
accuracy level.
\end{abstract}

\maketitle


\section{Experiment Goals and Motivation}

The PIBETA experiment\cite{pibeta} at the Paul Scherrer Institute
(PSI) is a comprehensive set of precision measurements of the rare
decays of the pion and muon.  The goals of the experiment's first
phase are:

\begin{itemize}

\item[(a)] To improve the experimental precision of the pion beta
decay rate, $\pi^+ \to \pi^0 e^+ \nu$ (known as $\pi_{e3}$, or
$\pi\beta$), from the present $\sim 4\,\%$ to $\sim 0.5\,\%$.  The
improved experimental precision will begin to approach the theoretical
accuracy in this process, and thus for the first time enable a
meaningful extraction of the CKM parameter $V_{ud}$ from a
non-baryonic process.

\item[(b)] To measure the branching ratio (BR) of the radiative decay
$\pi\to e\nu\gamma$ ($\pi_{e2}R$, or RPD), enabling a precise
determination of the pion form factor ratio $F_A/F_V$, and, hence, of
the pion polarizability.  Due to expanded phase space coverage of the
new measurement, we also aim to resolve the longstanding open question
of a nonzero tensor pion form factor.

\item[(c)] A necessary part of the above program is an extensive
measurement of the radiative muon decay rate, $\mu\to e \nu \bar{\nu}
\gamma$, with broad phase space coverage.  This new high-statistics
data sample is conducive to a precision search for non-\,$(V-A)$
admixtures in the weak Lagrangian.

\item[(d)] Both the $\pi\beta$ and the $\pi_{e2}R$ decays are
normalized to the $\pi\to e \nu$ (known as $\pi_{e2}$) decay rate.
The first phase of the experiment has, thus, produced a large sample
of $\pi_{e2}$ decay events.  The second phase of the PIBETA program
will seek to improve the $\pi_{e2}$ decay branching ratio precision
from the current $\sim 0.35\,\%$ to under 0.2\,\%, in order to provide
a precise test of lepton universality, and thus of certain possible
extensions to the Standard Model (SM).

\end{itemize}

Recent theoretical work\cite{jaus01,ciri02} has demonstrated low
theoretical uncertainties in extracting $V_{ud}$ from the pion beta
decay rate, i.e., a relative uncertainty of $5\times 10^{-4}$ or less,
providing further impetus for continued efforts in improving the
experimental accuracy of this process.

\section{Experimental Method}\label{sec:exp_met}

\begin{figure}[htb]
\hbox to\hsize{\hss
\includegraphics[width=\hsize]{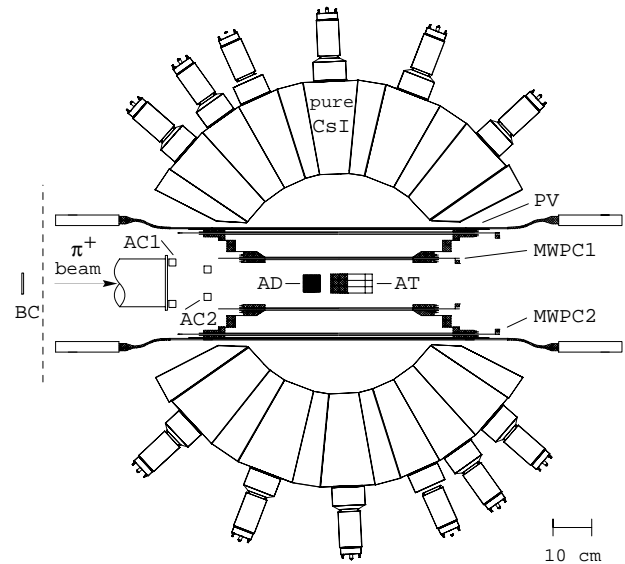}
\hss}
\caption{A schematic cross section of the PIBETA detector system.
Symbols denote: BC--thin upstream beam counter, AC1,2--active beam
collimators, AD--active degrader, AT--active target, MWPC1,2--thin
cylindrical wire chambers, PV--thin 20-segment plastic scintillator
barrel.  BC, AC1, AC2, AD and AT detectors are also made of plastic
scintillator.}
\label{fig:xsect}
\end{figure}

The $\pi$E1 beam line at PSI was tuned to deliver $\sim 10^6$
$\pi^+$/s with $p_\pi \simeq 113\,$MeV/c, that stop in a segmented
plastic scintillator target (AT).  The major detector systems are shown
in a schematic drawing in Fig.~\ref{fig:xsect}.  Energetic charged
decay products are tracked in a pair of thin concentric MWPC's and a
thin 20-segment plastic scintillator barrel detector (PV).  Both
neutral and charged particles deposit most (or all) of their energy in
a spherical electromagnetic shower calorimeter consisting of 240
elements made of pure CsI.  The CsI radial thickness, 22\,cm,
corresponds to 12$\,X_0$, and the calorimeter subtends a solid angle
of about 80\,\% of $4\pi\,$sr.

The basic principle of the measurement is to record all non-prompt
large-energy (above the $\mu \to e\nu\bar{\nu}$ endpoint)
electromagnetic shower pairs occurring in opposite detector
hemispheres (non-prompt two-arm events).  In addition, we record a
large prescaled sample of non-prompt single shower (one-arm) events.
Using these minimum-bias sets, we extract $\pi\beta$ and $\pi_{e2}$
event sets, using the latter for branching ratio normalization.  In a
stopped pion experiment these two channels have nearly the same
detector acceptance, and have much of the systematics in common.

A full complement of twelve fast analog triggers comprising all
relevant logic combinations of one- or two-arm, low- or high
calorimeter threshold , prompt and delayed (with respect to $\pi^+$
stop time), as well as a random and a three-arm trigger, were
implemented in order to obtain maximally comprehensive and unbiased
data samples.

\begin{figure}[htb]
\hbox to\hsize{\hss\includegraphics[width=\hsize]{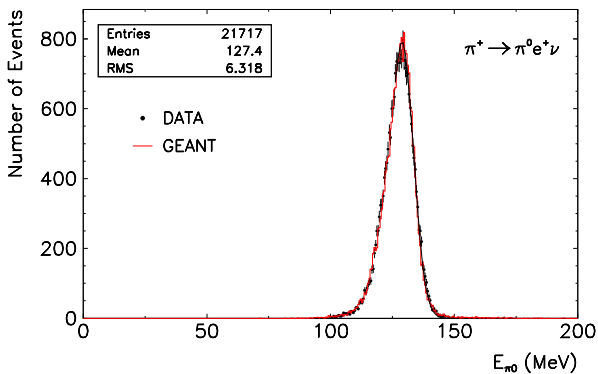}\hss}
\caption{$\pi^0$ energy spectrum for a subset of the measured
$\pi^+\to\pi^0e^+\nu$ decay data; solid curve: GEANT simulation.}
\label{fig:pb_en}
\end{figure}

\begin{figure}[htb]
\hbox to\hsize{\hss\includegraphics[width=\hsize]{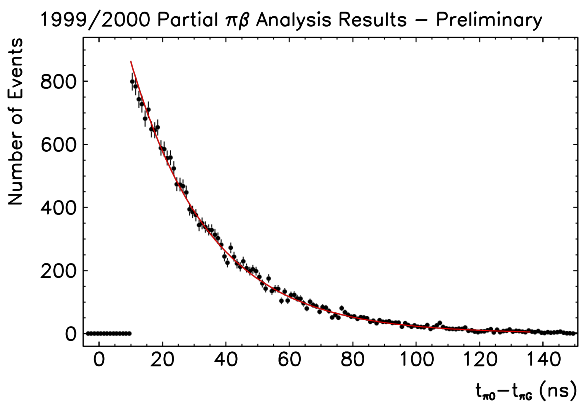}\hss}
\caption{Histogram of time differences between the beam pion stop and
the $\pi\beta$ decay events (dots); curve: pion lifetime exponential
curve.  A software cut at 10\,ns was applied.}
\label{fig:pb_tim}
\end{figure}

\begin{figure}[htb]
\hbox to\hsize{\hss\includegraphics[width=\hsize]{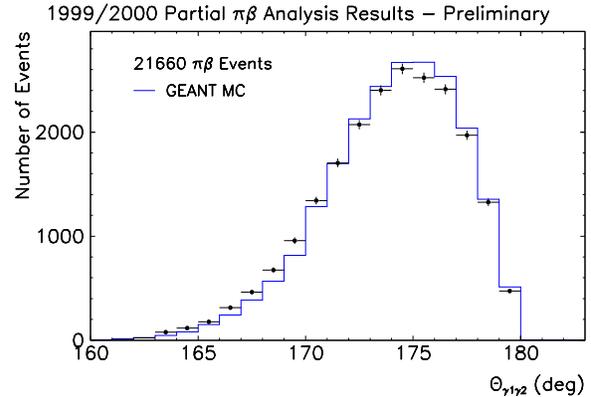}\hss}
\caption{Histogram of the measured $\gamma$-$\gamma$ opening angle
in pion beta decay events ($\pi^+\to\pi^0e^+\nu$) for a subset of
acquired data; solid curve: GEANT simulation.}
\label{fig:pb_th12}
\end{figure}

\begin{figure}[htb]
\hbox to\hsize{\hss\includegraphics[width=\hsize]{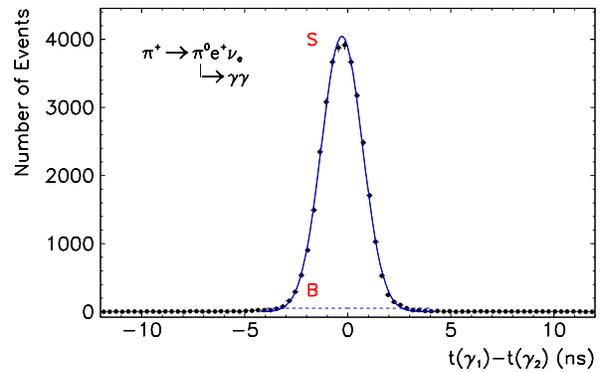}\hss}
\caption{Histogram of $\gamma$-$\gamma$ time differences for
the same set of $\pi\beta$ data events (dots); curve: fit.  Signal to
background ratio exceeds 250.}
\label{fig:pb_sn}
\end{figure}

The high quality of the PIBETA data is clearly demonstrated in the
histograms of the calorimeter energy and event timing (following the
$\pi^+$ stop time), as well as of the $\gamma-\gamma$ opening angle
and time difference for a subset of the recorded pion beta decay
events, shown in Figs.~\ref{fig:pb_en}--\ref{fig:pb_sn}.

In particular, the low level of accidental background is evident in
the $\gamma-\gamma$ relative timing histogram in Fig.~\ref{fig:pb_sn};
the peak to background ratio exceeds 250.  The histogram of recorded
$\gamma-\gamma$ opening angles for pion beta events, shown in
Fig.~\ref{fig:pb_th12}, provides possibly the most sensitive test of
the Monte Carlo simulation of the apparatus, and of the systematics
related to the geometry of the beam pion stopping distribution.  The
latter is the single largest contributor to the overall uncertainty in
the acceptance, and, hence, in the branching ratio.

\section{First Results: Pion Beta Decay}

The first phase of measurements took place in 1999, 2000 and 2001,
resulting in some 60,000 recorded pion beta events.  The figures of
Section~\ref{sec:exp_met} are based on a data subset acquired in 1999
and 2000.  Our current {\sl\bfseries preliminary working} result for
the pion beta decay branching ratio, extracted from the above
analysis, is
\begin{equation}
   BR \simeq 1.044 \pm 0.007{\rm (stat.)} \pm  0.009{\rm (syst.)} 
             \times 10^{-8}\ .       \label{eq:pb_br_exp}
\end{equation}
Our result is to be compared with the previous most accurate
measurement of McFarlane et al.\cite{McF85}: 
$$
   BR = 1.026 \pm 0.039 \times 10^{-8}\ ,
$$
as well as with the SM Prediction (Particle Data Group,
2002\cite{PDG02}): 
\begin{center}
\begin{tabular}{r@{\extracolsep{0.2em}}l}
   $BR =$ & $1.038 - 1.041 \times 10^{-8} \rm \quad (90\% C.L.)$ \\
        & $(1.005 - 1.008 \times 10^{-8} \rm \quad excl.\ rad.\ corr.)$\\
\end{tabular}
\end{center}
We see that even our working result strongly confirms the validity of
the CVC hypothesis and SM radiative corrections\cite{Mar86,jaus01,ciri02}.
Another interesting comparison is with the prediction based on the
most accurate evaluation of the CKM matrix element $V_{ud}$ using 
the CVC hypothesis and the results of measurements of superallowed
Fermi nuclear decays (Particle Data Group 2002\cite{PDG02}):
$$
   BR = 1.037 \pm 0.002 \times 10^{-8}\ .
$$
Thus, our current preliminary working result is in very good agreement
with the predictions of the Standard Model and the CVC hypothesis.
The quoted systematic uncertainties are being reduced as our analysis
progresses.  To put this result into broader perspective, we can
compare the central value of $V_{ud}$ extracted from our data with
that listed in PDG 2002\cite{PDG02}:
\begin{center}
\begin{tabular}{rl}
        {\rm PDG\ 2002:}  & $V_{ud} = 0.9734 (8)$,  \\
   {\rm PIBETA\ prelim:}  & $V_{ud} = 0.9771 (56)$. 
\end{tabular}
\end{center}

Table \ref{tab:pb_unc} summarizes the main sources of uncertainties
and gives their values both in the current analysis, and those that
are expected to be reached in a full analysis of the entire dataset
acquired to date.  We have temporarily enlarged the systematic
uncertainty quoted in Eq.~\ref{eq:pb_br_exp} pending a resolution of
the discrepancy found in the RPD channel and discussed in the
following section.

\begin{table}[ht]
\caption{Summary of the main sources of uncertainty in the extraction
of the pion beta decay branching ratio.  The column labeled
``current'' corresponds to the present analysis based on a portion of
the data taken.  }

\begin{center}

\begin{tabular}{rcc}

\hline\hline \\[-2ex]
  \multicolumn{1}{c}{Summary of}   
                     & \multicolumn{2}{c}{Dataset analyzed:}   \\
  \multicolumn{1}{c}{uncertainties (\%)}
                     &  partial$^*$ &    full        \\[0.5ex]
\hline \\[-1ex]
 \multicolumn{1}{l}{external:}                         \\
      pion lifetime           &  0.019   & 0.019       \\
      $BR(\pi \to e\nu)$      &  0.33    
                                & $\sim 0.1^\dagger$          \\
     $BR(\pi^0\to\gamma\gamma)$ &  0.032   & 0.032       \\[1ex]
 \multicolumn{1}{l}{internal:}                         \\
      $A(\pi\beta)/A(e\nu)$   &  0.5     & $<0.3$      \\
      $\Delta t(\gamma - e)$  &  0.03    & 0.03        \\
      E threshold             &  $<0.1$  & $<0.1$      \\[1ex]
 \multicolumn{1}{l}{statistical:}  &  0.7 & $\sim 0.4$ \\[1ex]
 \multicolumn{1}{l}{total:}   &  $\sim 0.9$  
                                       & $\lesssim 0.5$ \\[1ex]
\hline\hline\\[-1ex]

\multicolumn{3}{l}{$^*$ Subset of the 1999 and 2000 data.}    \\
\multicolumn{3}{l}{$^\dagger$ Requires a new measurement.}    \\[-2ex]

\end{tabular}

\label{tab:pb_unc}

\end{center}

\end{table}

\section{First Results: Radiative Pion Decay}

As was already pointed out, we have recorded a large data set of
radiative decays: $\pi^+\to e^+\nu\gamma$ and $\mu^+\to
e^+\nu\bar{\nu}\gamma$.  To date we have analyzed both pion and muon
radiative decays, though with more attention devoted to the former, as
it is an important physics background to other decays under study.
The pion radiative decay analysis has given us the most surprising
result to date, and has commanded significant effort on our part to
resolve the issue.

The different event triggers used in our experiment are sensitive to
three distinct regions in the RPD phase space:

\begin{itemize}

\item region A with $e^+$ and $\gamma$ emitted into opposite
hemispheres, each with energy exceeding that of the Michel edge ($E_M
\simeq 52\,$MeV), recorded in the main two-arm trigger,

\item region B with an energetic photon ($E_\gamma > E_M$), and
$E_{e+} \geqslant 20\,$MeV, recorded in the one-arm trigger, and

\item region C with an energetic positron ($E_{e+} > E_M$), and
$E_\gamma \geqslant 20\,$MeV, also recorded in the one-arm trigger.

\end{itemize}

The RPD data are of a similar quality to our $\pi\beta$ event set; due
to limited space we omit the details here, and direct the interested
reader to Ref.~\cite{pibeta} instead.

Together, the three regions overconstrain the Standard Model
parameters describing the decay, and thus allow us to examine possible
new information about the pion's hadronic structure, or non-(V$-$A)
interactions.  Appropriate analysis of these data is involved and
nuanced, requiring a longer presentation than is possible here.  We
therefore only summarize the salient results of our work in progress
on this pion decay channel.

Our analysis indicates a measurable departure from SM predictions.
Standard Model with the V$-$A electroweak sector requires only two
form factors, $F_A$ and $F_V$ to describe the so-called
structure-dependent amplitude in RPD.  The remainder of the decay
amplitude is accounted for by QED in the inner-bremsstrahlung (IB)
term.  The pion vector form factor is strongly constrained by the CVC
hypothesis, while existing data on the radiative pion decay (PDG
2002\cite{PDG02}) suggest that $F_A \simeq 0.5\,F_V$, yielding
$$
    F_V = 0.0259 \pm 0.0005 \ , \qquad {\rm and} \qquad
    F_A \simeq 0.012\ .
$$
Simultaneous as well as separate fits of our data in regions A, B and
C confirm the above ratio of $F_A/F_V \simeq 0.5$.  However, they show
a statistically significant deficit in RPD yield in one region of
phase space, for high $E_\gamma$ and lower $E_e$ (mostly in region B),
compared to predictions based on the above values of the pion form
factors.

A larger deficit in RPD yield, though less statistically significant
than our result due to far fewer events, was first reported by the
ISTRA collaboration\cite{Bol90a,Bol90b}.  This first observation was
interpreted by Poblaguev\cite{Pob90,Pob92} as indicative of the
presence of a tensor weak interaction in the pion, giving rise to a
nonzero tensor pion form factor $F_T \sim -6 \times 10^{-3}$.
Subsequently, Peter Herczeg\cite{Her94} found that the existing
experimental evidence on beta decays could not rule out a small
nonzero value of $F_T$ of this order of magnitude.  Tensor interaction
of this magnitude could only be explained by the existence of
leptoquarks.

\begin{figure}[ht]
\includegraphics[width=\hsize]{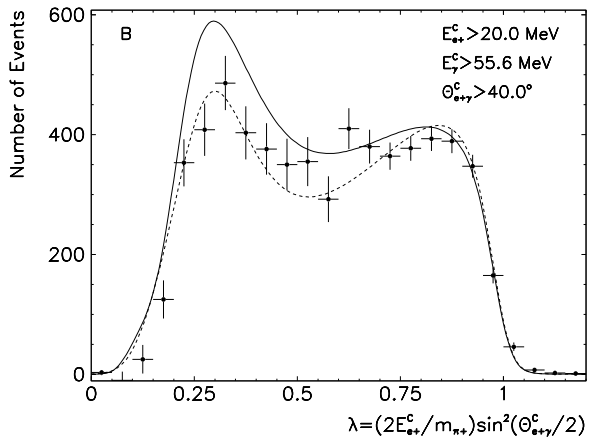}
\caption{\small Measured spectrum of the kinematic variable $\lambda =
(2E_{e+}/m_{\pi+})\sin^2(\theta_{e\gamma}/2)$ in $\pi^+ \to
e^+\nu\gamma$ decay for the kinematic region B, with limits noted in
the figure.  Solid curve: fit with the pion form factor $F_V$ fixed by
the CVC hypothesis, $F_T=0$, and $F_A$ free.  Dashed curve: as above,
but with $F_T$ also released to vary freely, resulting in $F_T =
-0.0016\,(3)$.  Work in progress.}
\label{fig:pienug:2a_l1}
\end{figure}

We illustrate our working results in Fig.~\ref{fig:pienug:2a_l1} which
shows a projected one-dimensional distribution of $\lambda$, a
convenient kinematic variable based on $E_e$ that ranges from 0 to 1
regardless of $E_\gamma$.  It is clear that for lower values of
$\lambda$ (and therefore of $E_e$), an SM fit with only $F_V, F_A \neq
0$ overestimates the experimental yield.  Adding a nonzero tensor form
factor of $F_T \sim -0.0016$ produces statistically significantly
better agreement with the data.  The fits are two-dimesional and
encompass all three kinematic regions, A, B, and C.  This work is in
progress, and the reuslts are subject to change---we are currently
refining the analysis as well as the fit strategies.

Taken at face value, this working result should not be interpreted as
an indication of the existence of a tensor weak interaction, i.e., of
leptoquarks.  Instead, if it holds up in our final analysis, it would
first suggest that the standard treatment of the RPD may not at this
time correctly incorporate all known SM physics.  Radiative
corrections seem to be a particularly good candidate for
reexamination.

\section{Conclusions}

We have extracted an experimental branching ratio for the pion beta
decay at the 1\,\% uncertainty level, and expect to reduce the
uncertainty by another factor of about two in the near future.  Our
result agrees with the CVC hypothesis and radiative corrections for
this process, and it opens the way for the first meaningful extraction
of the CKM parameter $V_{ud}$ from a non-baryonic process.

Our analysis of the $\pi\to e\nu\gamma$ decay confirms that $F_A/F_V
\simeq 0.5$, in agreement with the world average.  However, events
with a hard $\gamma$ and soft e$^+$ are not well described by standard
theory, requiring ``$F_T \neq 0$''.  A new theoretical look at this
decay is needed.  We can, though, rule out a large ``$F_T$'', as
reported in analyses of the ISTRA data.

The high statistics and broad coverage of our RPD data in principle
guarantee extraction of pion weak form factor values with
exceptionally low uncertainties.  However, it appears that there may
be significant theoretical uncertainties in the process of the form
factor extraction.  We hope that any remaining theoretical questions
will be resolved in the near future, eventually enabling the field to
make use of the full potential of the PIBETA data.

\end{document}